\documentstyle[twocolumn,aps,epsf]{revtex}
\begin{document}
\draft
%
%
%
%
\title{The Weak Localization Correction
to the Polarization and Persistent Currents in Mesoscopic Metal Rings}
\author{Peter Kopietz and Axel V\"{o}lker}
\address{
Institut f\"{u}r Theoretische Physik der Universit\"{a}t G\"{o}ttingen,
Bunsenstrasse 9, D-37073 G\"{o}ttingen, Germany}
\date{September 1, 1997}
\maketitle
\begin{abstract}
We re-examine the effect of electron-electron interactions
on the persistent current in mesoscopic metal rings
threaded by an Aharonov-Bohm flux.
The exchange contribution to the current
is shown to be determined by the
weak localization correction to the polarization.
We explicitly calculate the contribution from
exchange interactions with 
momentum transfers smaller than the inverse
elastic mean free path to the average current, and find that it
has the same order of magnitude as 
the {\it{canonical}} current  
without interactions. 

\end{abstract}
\narrowtext

The persistent current in mesoscopic metal rings threaded by a static magnetic
flux $\phi$ is a striking manifestation  of  
the quantum nature of charge transport in phase-coherent 
systems\cite{Hund38,Buttiker83}.
At the first sight one might expect that this effect
can be explained  within a model
of non-interacting electrons in a random potential.
Surprisingly, 
the seminal experiment by 
L\'{e}vy {\it{et al.}}\cite{Levy90}
revealed that in the diffusive regime the average current
is more than two orders of magnitude larger than
the theoretically predicted current for non-interacting electrons\cite{Schmid91}.
Therefore it seems inevitable to invoke
electron-electron interactions to explain the experiment\cite{Levy90}.
This was first attempted by
Ambegaokar and Eckern (AE)\cite{Ambegaokar90}, 
who calculated perturbatively the correction
to the average current to first order in the
screened Coulomb interaction.
They found an average current proportional to 
 $ \lambda_{\rm cb} {e v_F} \ell /{L}^2$, where 
$\lambda_{\rm cb}$ is a dimensionless measure
for some suitably averaged strength of the Coulomb interaction
at short wavelengths,
$v_F$ is the Fermi velocity, 
$\ell$ is the elastic mean free path, and
$L$ is the circumference of the ring. 
The precise value of 
$\lambda_{\rm cb}$ is difficult to estimate, because it is
dominated by the non-universal short-wavelength part of the Coulomb interaction.
Using \cite{Eckern91} $\lambda_{\rm cb} \approx 0.06$,
the theoretical current is still a factor of $5$ too small to explain
the experiment.

More recently one of us 
has argued that the classical (Hartree) contribution
to the persistent current is strongly enhanced
due to the long-range nature of the bare Coulomb
interaction\cite{Kopietz93}.
Although the arguments
put forward in Ref.\cite{Kopietz93} have been criticized\cite{Vignale94},
we believe that the criticism
did not properly take the essentially non-perturbative nature
of these arguments into account\cite{Kopietz94}.
However, this is not the subject of this note.

Here we would like to re-examine the exchange (Fock) contribution
to the average current.
It is well known\cite{Altshuler85} that for disordered metals
singular vertex corrections 
involving so-called diffusons
strongly enhance the effective exchange interaction at
small wave-vectors 
$|{\bf{q}} | 
\raisebox{-0.5ex}{$\; \stackrel{<}{\sim} \;$} 2 \pi /\ell$.
This effect has been ignored by AE\cite{Ambegaokar90}, who
focused on the short-wavelength part of the interaction.
The possible relevance of these diffusive vertex corrections
for persistent currents
has been pointed out some time ago by
B\'{e}al-Monod and Montambaux\cite{Beal92}.
However, they also demonstrated that there exists
an overall cancellation of the leading infrared singularities.
Up until now
a quantitative calculation of the persistent current
due to exchange interactions with small momentum transfers
has not been performed.
 In this work we shall present a
simple solution of this problem.

It is instructive to obtain the
(grand canonical) persistent
current $I = -c \partial \Omega ( \phi ) / \partial \phi$
from the Grassmannian functional integral representation
of the ratio
${\cal{Z}}/{\cal{Z}}_0 = e^{- ( \Omega - \Omega_0)/T}$ of the partition functions
with and without interactions. Here $T$ is the temperature.
Decoupling the Coulomb interaction
by means of a Hubbard-Stratonovich field $V$ and 
integrating over the Grassmann fields, we obtain after the
usual transformations\cite{Kopietz97}
 ${\cal{Z}} /{\cal{Z}}_0 = \int {\cal{D}} \{ V \} e^{ - S_{\rm eff} \{ V \}}$,
with the effective action\cite{footnote1}
 \begin{equation}
 S_{\rm eff} \{ V \} = \frac{T}{2} \sum_{q} f^{-1}_{\bf{q}} V_{-q} V_q
 -  2 {\rm Tr} \ln \left[ 1 - \hat{G}_0 \hat{V} \right]
 \;  ,
 \label{eq:Seffdef}
 \end{equation}
where the factor of $2$ in front of the trace is due to the
spin degeneracy, and
$f_{\bf{q}}$ is the Fourier transform of the
Coulomb potential\cite{footnote2}.
$\hat{V}$  and $\hat{G}_0$ are infinite matrices in
frequency and wave-vector space, with matrix elements
$[\hat{V}]_{k k^{\prime} } = i T V_{k-k^{\prime}} $
and\cite{footnote1}
 \begin{equation}
 [\hat{G}_0^{-1}]_{k k^{\prime}} = \delta_{k k^{\prime}} [ i \tilde{\omega}_n
 - \xi_{\bf{k}} ] - \delta_{ n  n^{\prime} } U_{{\bf{k}} - {\bf{k}}^{\prime}} 
 \label{eq:G0hatdef}
 \;  .
 \end{equation}
Here $U_{\bf{q}}$ is the
Fourier transform of the disorder potential (assumed
to be Gaussian white noise), and
$\xi_{\bf{k}} = {\bf{k}}^2 / ( 2 m ) - \mu$,
where $m$ is the (effective) mass of the electrons, and $\mu$ is the
chemical potential.
Identifying the position along the circumference
of the ring with the $x$-coordinate, the flux-dependence 
of $\Omega ( \phi )$ is due to the quantization
$k_x = \frac{2 \pi}{L} ( n + \phi / \phi_0 ) $, $n = 0, \pm 1, \ldots$ of the
$x$-component of the wave-vector. 

For a calculation to first order
in the RPA (random phase approximation) screened interaction 
it is sufficient to expand the logarithm in
Eq.(\ref{eq:Seffdef}) to second order, 
 \begin{eqnarray}
 S_{\rm eff} \{ V \} & \approx &
 i \sum_{{q}} N_0 ( q ) V_{- {q}} 
 \nonumber 
 \\
 &  & \hspace{-15mm} +
 \frac{T}{2 } \sum_{q q^{\prime}} \left[ \delta_{qq^{\prime}} f^{-1}_{\bf{q}}  +
 \Pi_0 ( q , q^{\prime} ) \right]
 V_{-q } V_{q^{\prime}} + \ldots
 \label{eq:Seffexpansion}
 \; \; ,
 \end{eqnarray}
where $N_0 ( {{q}} ) = 2 T \sum_k [ \hat{G}_0 ]_{k +q, k}$
and
 \begin{equation}
 \Pi_0 ( q , q^{\prime} ) = - 2 T \sum_{k k^{\prime}}
 [\hat{G}_0]_{k+q,k^{\prime} + q^{\prime}} [\hat{G}_0]_{k^{\prime}k}
 \label{eq:pol}
 \;  .
 \end{equation}
Note that $[\hat{G}_0]_{k k^{\prime}}$  
in Eq.(\ref{eq:G0hatdef})
is proportional to $\delta_{n  n^{\prime}}$, so that
$N_0 (q ) = \delta_{n0} N_0 ( {\bf{q}} )$ 
can be identified with
the spatial Fourier component of the density for a given realization
of the disorder, and
$ \Pi_0 ( q , q^{\prime} ) = \delta_{n  n^{\prime}} 
\Pi_0 ( {\bf{q}} , {\bf{q}}^{\prime} , i \omega_n )$
is the non-interacting polarization.
The linear term in Eq.(\ref{eq:Seffexpansion}) can be eliminated by means of
the shift-transformation
$V_q \rightarrow V_q - i T^{-1} \sum_{q^{\prime}} f^{\rm RPA}_{ q  q^{\prime} } 
N_0 ( q^{\prime } )$, where $f^{\rm RPA}_{  q  q^{\prime}}$ is the
{\it{inverse}} of the infinite matrix with elements
$\delta_{q q^{\prime}} f^{-1}_{\bf{q}} + \Pi_0 ( q , q^{\prime} )$. 
To calculate the persistent current it is convenient 
to take the derivative of $ \ln ( {\cal{Z}} / {\cal{Z}}_0 )$
with respect to $\phi$ {\it{before}} performing the Gaussian
integration over the auxiliary field $V$. 
In this way we obtain
$I = I_{0} + I_{ H} + I_{ F}$, where
$I_{0}$ is the non-interacting persistent current, and 
 \begin{eqnarray}
 I_{ H} & = & \frac{-c}{2} \frac{ \partial}{\partial \phi }
 \sum_{ {\bf{q}}  {\bf{q}}^{\prime} } 
 f^{\rm RPA}_{ {\bf{q}} {\bf{q}}^{\prime} }
 N_0 ( - {\bf{q}} ) N_0 ( {\bf{q}}^{\prime} )
 \label{eq:IH}
 \;  ,
 \\
 I_{ F} & = & \frac{-c}{2} 
 \sum_{ {\bf{q}}  {\bf{q}}^{\prime} } T \sum_{n}
 f^{\rm RPA}_{ {\bf{q}} i \omega_n , {\bf{q}}^{\prime} i {\omega}_n} 
 \frac{ \partial}{\partial \phi }
 \Pi_0 ( {\bf{q}} , {\bf{q}}^{\prime} , i \omega_n )
 \;  .
 \label{eq:IF}
 \end{eqnarray}
Note that the Hartree contribution $I_H$
involves only the static part 
$ f^{\rm RPA}_{ {\bf{q}} {\bf{q}}^{\prime} } \equiv 
 f^{\rm RPA}_{ {\bf{q}} 0, {\bf{q}}^{\prime} 0 }$
of the screened interaction.

\begin{figure}
\epsfysize5cm 
\hspace{5mm}
\epsfbox{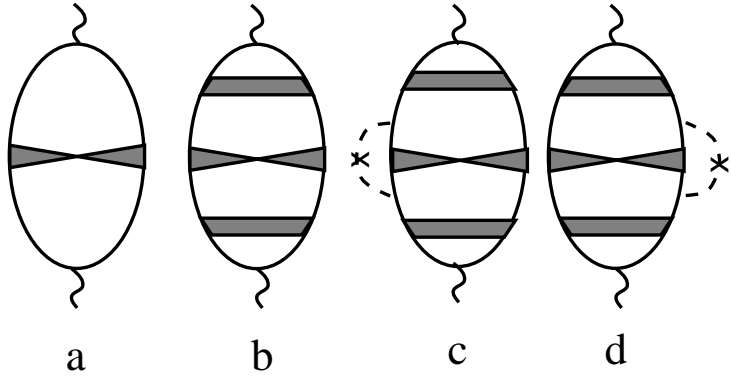}
\vspace{5mm}
\caption{
Diagrams contributing to
$\partial \overline{\Pi}_0 ( {\bf{q}} , i \omega )/ \partial \phi$.
Retaining only (a) reproduces the result of AE.
The shaded triangles represent the Cooperon, the
solid lines are average non-interacting Green's functions, and the
small wavy lines denote density vertices.
(b) Vertex correction with two additional diffusons
(represented by shaded stripes).
For small ${\bf{q}}$ and $\omega$ this diagram 
is much larger than (a).
However, to leading order in $ ( k_F \ell )^{-1}$ 
(b) is cancelled by (c) and (d).
Here the dashed lines denote the 
impurity average $\overline{U_{\bf{q}} U_{\bf{-q}}}$.
}
\label{fig:pol}
\end{figure}

Let us now focus on the average Fock current $\overline{I}_{ F}$
in the diffusive regime, where
the over-bar denotes the disorder average.
To leading order in the small parameter $  (k_F \ell)^{-1} $ 
(where $k_F$ is the Fermi wave-vector)
we may factorize the average, so that
we need to know
the flux-derivative of the average
polarization $\partial \overline{\Pi}_0 ( {\bf{q}} , i \omega ) / \partial \phi$.
Note that after averaging the polarization is diagonal in 
wave-vector space. It is easy to show\cite{Voelker97a}
that the result of AE\cite{Ambegaokar90} can be 
reproduced if one
retains only diagram (a) in Fig.\ref{fig:pol}. 
For $| {\bf{q}} | \ll k_F$ and $| \omega | \tau \ll 1$
(where $\tau = \ell / v_F$ is the elastic lifetime)
this diagram yields  
%
%
%
%
%
%
%
%
 \begin{equation}
  \frac{\partial}{\partial \phi} 
 \overline{\Pi}_0 ( {\bf{q}} , i \omega )_{\rm AE} 
 = \Delta^{-1} {| \omega | \tau } \frac{\partial}{\partial \phi} 
 g_{\rm WL}  ( i \omega , \phi )
 \label{eq:PiphiAE}
 \;  ,
 \end{equation}
where $\Delta$ is the average level-spacing
at the Fermi energy.
$g_{\rm WL} ( i \omega, \phi)$ is the well known weak-localization correction
to the dimensionless average conductivity $\overline{\sigma} ( i \omega
)$ in units of the
corresponding Drude conductivity $\sigma_0$,
 \begin{equation}
 \overline{\sigma} ( i \omega ) = \sigma_0 [ 1 + g_{\rm WL}  ( i
 \omega , \phi )]
 \; ,
 \label{eq:sigmadrudewl}
 \end{equation}
 \begin{equation}
 g_{\rm WL} ( i \omega , \phi ) = - \frac{2 \Delta}{\pi} {\sum_{ 
  {\bf{Q}}  }}^{\prime}
 \frac{ 1}{ {\cal{D}}_0  {\bf{Q}}^2 + | {\omega} | + {\cal{D}}_0 / L_{\varphi}^2 }
 \; ,
 \label{eq:gWL}
 \end{equation}
where ${\cal{D}}_0 = v_F \ell / 3$ is the classical
diffusion coefficient.
Note that ${\cal{D}}_0$ and $\sigma_0$ are related
via the classical Einstein relation
$\sigma_0 = ( \Delta {\cal{V}})^{-1} e^2 \sigma_0$, where
${\cal{V}}$ is the volume of the system.
The prime in Eq.(\ref{eq:gWL}) means that
the sum is restricted to 
$| {\bf{Q}} | 
\raisebox{-0.5ex}{$\; \stackrel{<}{\sim} \;$} 2 \pi /\ell$.
Here the $x$-component of ${\bf{Q}}$
is quantized according to ${Q}_x = \frac{2 \pi}{L} (n + 2 \phi/ \phi_0 )$,
and $L_{\varphi}$ is the dephasing length\cite{Altshuler85}.

Clearly, the approximation (\ref{eq:PiphiAE})
cannot be correct for $| {\bf{q}} | \ell \ll 1$,
because then the density vertices in Fig.\ref{fig:pol}(a)
can be dressed by singular diffuson corrections, as shown
in Figs.\ref{fig:pol}(b)--(d). 
As a consequence, this regime requires a special treatment.
This has already been noticed by B\'{e}al-Monod and Montambaux\cite{Beal92}.
Unfortunately, the sum of diagrams (b)--(d)
in Fig.\ref{fig:pol} cancels to leading order
in $( k_F \ell )^{-1}$\cite{Beal92}, so that at the first sight
it seems that the calculation of the leading
non-vanishing behavior of
$\partial \overline{\Pi}_0 ( {\bf{q}} , i \omega ) / \partial \phi$
requires rather complicated mathematical
manipulations.
However, there is a simple way to avoid this technical complication:
As shown in Ref.\cite{Vollhardt80},
current conservation implies that 
for small wave-vectors and frequencies the polarization
is of the form
 \begin{equation}
 \overline{\Pi}_0 ( {\bf{q}} , i \omega ) = \Delta^{-1}
 \frac{ {\cal{D}}  ( i \omega ) {\bf{q}}^2 }{ {\cal{D}} ( i \omega
 ){\bf{q}}^2 + 
 | \omega | } 
 \label{eq:Pisigma0}
 \; ,
 \end{equation}
where ${\cal{D}} ( i \omega)$ is the generalized frequency-dependent diffusion
coefficient, which is related to the
frequency-dependent average conductivity 
$\overline{\sigma} ( i \omega )$ via\cite{Vollhardt80}
 \begin{equation}
 \frac{ {\cal{D}} ( i \omega )}{{\cal{D}}_0}
 = \frac{ \overline{\sigma} ( i \omega ) }{\sigma_0}
 \; .
 \label{eq:sigmaD}
 \end{equation}
Combining Eq.(\ref{eq:sigmaD}) with
Eqs.(\ref{eq:sigmadrudewl}) and (\ref{eq:gWL}), we conclude
 \begin{equation}
 \frac{ \partial}{\partial \phi} {\cal{D}} ( i \omega)
 = {\cal{D}}_0  
 \frac{ \partial}{\partial \phi}
 g_{\rm WL} ( i \omega , \phi )
 \; .
 \label{eq:sigmaDD}
 \end{equation}
From Eqs.(\ref{eq:Pisigma0}) and (\ref{eq:sigmaDD}) it is now obvious
that the flux-dependence of the average polarization 
can be expressed in terms of
the weak localization correction (\ref{eq:gWL}) to the dynamic conductivity.
After a straightforward calculation we finally obtain
 \begin{eqnarray}
  \frac{\partial}{\partial \phi} 
 \overline{\Pi}_0 ( {\bf{q}} , i \omega ) 
 & = & \Delta^{-1}  
 \frac{ | \omega | {\cal{D}}_0  {\bf{q}}^2 }{ 
 \left[ {\cal{D}}_0 {\bf{q}}^2 [ 1 + g_{\rm WL} ( i \omega , \phi )]  
 + | \omega | \right]^2} 
 \nonumber
 \\
 & \times &
 \frac{\partial}{\partial \phi} g_{\rm WL}  ( i \omega , \phi )
 \label{eq:Piphires}
 \;  .
 \end{eqnarray}
This expression, which is one of the main results of this
work, is  valid in the regime
$| {\bf{q}} | \raisebox{-0.5ex}{$\; \stackrel{<}{\sim} \;$} 2 \pi /\ell$ 
and 
$\Delta \ll | \omega | \raisebox{-0.5ex}{$\; \stackrel{<}{\sim} \;$}
\tau^{-1}$.
For
$| \omega | \raisebox{-0.5ex}{$\; \stackrel{<}{\sim} \;$} \Delta$
more sophisticated methods are necessary to obtain the
weak localization correction to the  polarization\cite{Efetov95}.
For our purpose Eq.(\ref{eq:Piphires}) is sufficient, because
the average of Eq.(\ref{eq:IF}) is dominated by
frequencies in the range
$\Delta \ll | \omega | \raisebox{-0.5ex}{$\; \stackrel{<}{\sim} \;$}
E_c$, where $E_c = {\cal{D}}_0 / L^2$ is the Thouless energy. 
Note that for $| {\bf{q}} | \ell = O (1)$ Eq.(\ref{eq:Piphires}) smoothly matches
the short wavelength result (\ref{eq:PiphiAE}).
In fact, using ${\cal{D}}_0 = v_F \ell /3$, we see that
for $| g_{\rm WL} | \ll 1$
both expressions are identical 
at $| {\bf{q}}| \ell  = \sqrt{3}$.
On the other hand, in the regime $|{\bf{q}} | \ell 
\ll 1$ the long-wavelength result
(\ref{eq:Piphires}) is a factor of
$ ( {\bf{q}} \ell )^{-2}$
larger than Eq.(\ref{eq:PiphiAE}).

To see whether this infrared enhancement 
is sufficient to lead to a significant exchange contribution
to the persistent current, we substitute Eq.(\ref{eq:Piphires})
into Eq.(\ref{eq:IF}). Then we obtain
for the long-wavelength Fock contribution to the average current
 \begin{eqnarray}
 \overline{I}_{ F}^{\rm long} & =  &  
 - \frac{c}{2 }  T \sum_n
 {\sum_{\bf{q}}}^{\prime}
 \frac{ f_{\bf{q}} \frac{\partial}{\partial \phi}
 \overline{\Pi}_0 ( {\bf{q}} , i \omega_n )}{ 1+
 f_{\bf{q}} \overline{\Pi}_0 ( {\bf{q}} , i \omega_n ) }
 \nonumber
 \\
 &  \approx &
 - \frac{c}{2 } 
 T \sum_n
 \left[
 {\sum_{\bf{q}}}^{\prime}
  \frac{ | \omega_n |}{ {\cal{D}}_0  {\bf{q}}^2
 [ 1 + g_{\rm WL} ( i \omega_n , \phi ) ]  +
 | \omega_n | } \right]
 \nonumber
 \\
 & & \times
 \frac{
 \frac{\partial}{\partial \phi} g_{\rm WL} ( i \omega_n , \phi )}{ 
 1 +
 g_{\rm WL} ( i \omega_n , \phi )}
 \label{eq:IFsum}
 \;  ,
 \end{eqnarray}
where the
second line is valid in the experimentally relevant limit 
 $ | f_{\bf{q}}  \overline{\Pi}_0 ( {\bf{q}} , i \omega_n ) | \gg 1$.
Note that the current (\ref{eq:IFsum})
increases with the strength of the disorder, because
the (negative) weak-localization correction  $ g_{\rm WL}$
in the denominator of Eq.(\ref{eq:IFsum})
becomes more and more important with increasing disorder, thus
reducing the screening.
Of course, our calculation is only
controlled for $|g_{\rm WL}| \ll 1$,  
so that from now on this correction will be ignored.
For simplicity let us assume that the ring is sufficiently thin,
such that only the motion along the circumference is diffusive.
In this case the ${\bf{q}}$-sum in the square brace of
Eq.(\ref{eq:IFsum}) can be replaced by a one-dimensional integral,
which for $L \gg \ell$ is independent of the ultraviolet cutoff.
We obtain
 \begin{equation}
 \overline{I}_F^{\rm long} = - \frac{c}{4} T \sum_n
 \left[  \frac{ | \omega_n | L^2}{  {\cal{D}}_0 } \right]^{1/2}
 \frac{\partial}{\partial \phi} g_{\rm WL} ( i \omega_n , \phi )
 \; .
 \end{equation}
For  $T \rightarrow 0$ and $L_{\varphi} \gg L$ the
summation over the Matsubara frequencies can be performed exactly,
and we finally arrive at
 \begin{equation}
 \overline{I}_F^{\rm long} = - \frac{ e v_F}{L} \frac{ \pi}{ ( k_F L_{\bot} )^2}
 f ( \phi)
 \; ,
 \end{equation}
where $L_{\bot}$ is the transverse thickness of the ring, and 
 \begin{eqnarray}
 f ( \phi) & = &
 \frac{2}{\pi} \sum_{m = 1}^{\infty} \frac{ \sin ( 4 \pi m \phi /
 \phi_0)}{m}
 \nonumber
 \\
 & = & 1 - 4 \phi / \phi_0 
 \; \; \; , \; \; \; \mbox{for $ 0 < \phi / \phi_0 < 1/2 $}
 \label{eq:fres}
 \; .
 \end{eqnarray}
Surprisingly, this current has the same order of magnitude
as the non-interacting current at constant particle number\cite{Schmid91}.
In the experimentally relevant parameter regime\cite{Levy90}, this
current is smaller than the current due to the short wavelength
part of the Coulomb interaction calculated by
AE\cite{Ambegaokar90}.

In summary, we have presented a quantitative calculation of
the long-wavelength exchange contribution to
the average persistent current in mesoscopic
metal rings.
The current has the same order of magnitude as the
{\it{non-interacting}} persistent current at constant
particle number.
We have also calculated the leading weak localization
correction to the average polarization $\overline{\Pi}_0 ( {\bf{q}} , \omega)$ 
in the regime
$| {\bf{q}} | \raisebox{-0.5ex}{$\; \stackrel{<}{\sim} \;$} 2
\pi/\ell$ and $\Delta  \raisebox{-0.5ex}{$\; \stackrel{<}{\sim} \;$} 
|\omega | \ll \tau^{-1}$, effectively taking
singular diffuson corrections to the density vertices into account.

This work was supported by the
Deutsche Forschungsgemeinschaft (SFB 345),
and by the
ISI Foundation (ESPRIT 8050 Small Structures).
The work of 
PK was partially carried out at the Villa Gualino, Torino.

%

%
%
%
%


\begin{thebibliography}{99}
\bibitem{footnote2}
In a finite system the Fourier transform $f_{\bf{q}}$ of the Coulomb potential
does not have any singularities. 
It is then convenient to normalize $f_{\bf{q}}$ 
in Eq.(\ref{eq:Seffdef}) such that
$f_{\bf{q}} \sim e^{2} /C_0$ for ${\bf{q}} \rightarrow 0$, where
$C_0$ is the classical capacitance\cite{Kopietz93}.
Similarly, the polarization is normalized such that to leading order
$\overline{\Pi}_0 ( {\bf{q}} , 0 ) \sim \Delta^{-1}$.
\end{thebibliography}
\end{document}